# HYBRID THREATS AS AN EXOGENOUS ECONOMIC SHOCK

**Shteryo S. Nozharov, Head Assist. Prof., PhD**[1]
*University of National and World Economy – Sofia,*
*Department of Economics*

**Abstract:** The aim of this study is to contribute to the theory of exogenous economic shocks and their equivalents in an attempt to explain business cycle fluctuations, which still do not have a clear explanation. To this end the author has developed an econometric model based on a regression analysis. Another objective is to tackle the issue of hybrid threats, which have not yet been subjected to a cross-disciplinary research. These were reviewed in terms of their economic characteristics in order to complement research in the fields of defence and security.

\* \* \*

### Introduction

The beginning of the 21st century was marked by economic, environmental, terrorist, technological, migration and other challenges and crises. They all act as shocks to social, political and economic systems at national, regional and global level. The processes of integration and globalization have increased the mobility of people, goods, capitals, ideas, and information. They also resulted in clashes of cultures, religions and nationalist doctrines.

Economic crises, such as the one that occurred in 2008, are becoming increasingly difficult to predict because economic processes and crises can no longer be explained only with endogenous factors intrinsic to the economic system. At the same time, economic theory has never set a limit to the number of possible causes of exogenous shocks. Not all environmental, technological, and political causes have been identified and studied comprehensively. This raises some pertinent questions, such as "Can we predict the next global economic crisis based on our current knowledge on economic processes" and "Can we ensure continuous economic growth and development."

---

[1] E-mail: nozharov@unwe.bg

The aim of this study is to investigate hybrid threats as a potential exogenous economic shock which distorts the normal business cycle fluctuations.

On the one hand, there are almost no scientific publications on the economic aspects of hybrid threats except terrorism. Thus the study would provide a new perspective and outline another potential cause for exogenous economic shocks. Therefore, the primary objective of the study is to improve crisis forecasting and business cycle management as well as contribute to the economic theory in this field.

On the other hand, the number of interdisciplinary research studies on hybrid wars and attacks is quite insufficient. Thus, both the developed economies and the economies in transition cannot adapt to a future of conflicts of unforeseeable intensity and with unknown adversaries that undermine the existing socio-political systems. In this regard, the second objective of the study is to apply the economic approach to expand the theoretical knowledge on forecasting and management of hybrid attacks and wars.

The expected result of the study is an econometric model.

One of the limitations of the study is that the econometric model can be tested with the available data for Bulgaria only.

## 1. Definitions and literature review

The literature review could not identify any scientific publications on the effects of hybrid threats on the business cycle nor any systematic studies on the economic aspects of hybrid wars. Therefore, for the purposes of the literary analysis as well as the overall study, some basic definitions will be discussed.

For the purposes of this study we shall adopt the definition provided in the seminal publication of Hoffman (2009):

Hybrid threats incorporate a range of unique combinational threats specifically designed to target the attacked country's vulnerabilities. The instigator of a hybrid warfare is a state or a group of states that choose a non-conventional approach to warfare. The adversary selects from the whole menu of tactics and technologies and blends them in innovative ways to meet their own strategic culture, geography, and aims. Some of the methods may include a decentralized planning, cyberattacks, and nonstate actors (religious, criminal or separatist groups). What sets them apart from conventional terrorist groups is that they follow a unified strategy and are generally operationally and tactically directed and coordinated by a common unit to achieve synergistic effects of the high-tech capabilities and the resources of the instigator. The ultimate goal is no different from the goal pursued through conventional wars.

This definition is supported by Lerer and Amram-Katz (2011), Reeves and Barnsby (2013), Johnson and MacKay (2015).

The research which is closest to the subject matter of this study was reported by Guriev and Melnikov (2016) in their paper entitled "War, Inflation, and Social Capital". It is the only economic analysis with reference to the term "hybrid war". The aim of their research was to measure the impact of economic shocks and proximity to war on social capital using an economic approach. However, it differs in several ways from this study. First of all, the focus of their analysis is the attacking country rather than the attacked state. Secondly, although the term "hybrid war" is mentioned several times in the paper, it does refer to the final phase of a military operation involving the use of heavy weapons. Thirdly, the focus of the analysis is on social capital rather than the business cycle or the overall state of the economic system. In other words, they investigate the social response and support for the hybrid warfare operations that resulted in political and social tensions in the aggressor country.

Unlike the research conducted by Guriev and Melnikov (2016), this study focuses on the attacked country as well as on the initial hybrid warfare phases, which are characterized with low intensity and are therefore difficult to identify due to the lack of an actual armed conflict.

Other publications taken into consideration deal with the economic aspects of civil war and separatism (Chen, Loayza and Reynal-Querol, 2008; Blattman and Miguel, 2010; Cederman et al., 2015). They investigate the causality between civil war and poverty, religion, or education. They assume that the specific causes of civil war may be the indivisibility of desired goods (religious artefacts, sacred sites, cultural relics), asymmetric information available to the adversaries, flaws in the institutional framework (e.g. regulatory protectionism for the ruling elite or ethnic group, non-existent or corrupt voting system), population density, and internal geography. Methodologically, these studies are based on comparative analyses of economic and social parameters before and after armed conflicts taking into account the costs incurred during the conflict and adversaries' access to financing to estimate the possibility and the time needed for economic recovery after the end of the armed conflict. The costs include military spending, social costs in terms of mortality, disability, and infectious diseases, etc. and the demerits related to institutions, property rights, transaction costs, etc.

Contrary to these authors, this study is based on the underlying assumption that the causes of such conflicts are not endogenous for the attacked country but exogenous and geostrategic. Therefore, the correlation between the outbreak of a conflict and level of poverty or education in the attacked country is not relevant.

External instigation shows that the religious or ethnic fragmentation of the attacked country cannot be the leading cause of an active conflict. According to the definition of hybrid warfare adopted for this study, such conflicts do not necessarily involve the use of weapons and manslaughter and may be confined to cyberattacks, media misinformation, foreign electoral interventions, etc. Therefore, they may not result in costs incurred for the purchase and production of new weapons. This methodologically excludes the accounting of costs associated with casualties, injuries and diseases, damages to infrastructure, rearmament, etc. This is why the models used in the second group of publications and based on measuring the economic aspects of civil war and separatism are not applicable to hybrid threats and hybrid warfare and a different model must be used got the purposes of this study.

Still other publications deal with the economic aspects of terrorism (Abadie and Gardeazabal, 2003; Becker and Rubinstein, 2011; Eckstein and Tsiddon, 2004; Sandler and Enders, 2008; Meierrieks and Gries, 2013) defined as the deliberate use of violence or threat on an unspecified number of persons and the destruction of property, often with collateral murder of civilians who are not aware that they are being targeted by terrorists. The aim of terrorism is to exert pressure on the government in order to achieve the political or territorial claims of certain ethnic, social or religious groups mainly by generating mass fear and insecurity. Terrorism can be domestic or transnational.

There are several categories of costs associated with terrorism. The first group includes costs associated with the life and health of the victims of terrorist actions (death, disability, lost income for the victims and their relatives, loss of labour for the employers of the victims, healthcare and social services costs). The second category includes costs associated with the investment uncertainty created by terrorism (decrease of the volume of FDI, loss of infrastructure and factors of production, stock market uncertainty, decrease of the volume of savings, higher insurance and social security costs). The third category includes costs associated with increase of government spending on security and public order (additional police forces and intelligence operations in the target areas, higher salaries of public administrators in these regions) and the related reduction of government spending for other, more productive public services. According to these publications, all these costs affect economic growth by reducing the labour force, savings, and investments and raising the security costs in many profitable sectors, such as air transport and tourism.

In contrast to these publications, the present study is based on the following assumptions. Terrorism is one of the methods that can be used for hybrid warfare. Hybrid warfare is a military strategy adopted by one or several countries which use a combination of different methods. In addition to terrorism, the aggressor uses high-tech and other means of attack, such as media disinformation, cyber-attacks on the financial system, generation of targeted migration waves, etc. (Bachmann and Gunneriusson, 2015). Therefore, the two categories of terrorism commonly described in the scientific (domestic and transnational) must be augmented with a third category - "terrorism used as a hybrid warfare method". The existing scientific publications discuss terrorism as a detached phenomenon rather than a tool used in hybrid warfare. Terrorism may not necessarily be used as a hybrid warfare method. When the aggressor has substantial resources and estimates that the adversary has sufficient other vulnerabilities, he may limit himself to other means (cyber-attacks on the financial system, media disinformation) and do not resort to terrorism.

## 2. Economic analysis of the relationship between economic growth and hybrid warfare

Defining the variables for the model. Arnold's (2012) article is considered the seminal classification of the so-called "fourth generation warfare" (4GW) used to describe conflicts in the 21st century. In this type of conflict an adversary with limited resources can defeat a more powerful one. The lack of hierarchical authority and formal structure of the 4GW enemy, the low intensity of its operations and its lack of identification as a military formation make such conflicts difficult to recognize as wars by the international humanitarian law. This, in turn, does not allow the attacked country to defend using military force and prove the identification of its enemy. 4GW targets not only the military, but also the cultural, legal, economic and political systems of the enemy. Thus, the strengths of the attacked country are undermined and it will not be able to defend itself effectively.

Arnold (2012) classifies conflicts into three main groups: *regular warfare*, *irregular warfare* and *cosmopolitan action*. Regarding the 4GW especially interesting is the the second group, i.e. the *irregular warfare* methods, which include *insurrection*, *insurgency*, *guerrilla warfare*, complex irregular warfare, *advanced irregular warfare*, *compound warfare, hybrid warfare*, *criminal warfare and insurgency*, and *terrorism*. Arnold's (2012) classification allows us to analyse various combinations of hybrid warfare methods. Due to the limited scope of this study, it will not focus on a single method but on the determinants that apply to all hybrid warfare methods.

The analysis is based on the linear regression method. This particular method is chosen because of its easy application to pilot quantitative studies, for which when there are no established econometric models as well as the possibility for its approbation and verification of the results by other researchers.

The independent (factor) variables in the regression analysis are any unusual levels in the attacked country of:
1. **(A)** *public corruption crimes*;
2. **(B)** *cybercrimes*;
3. **(C)** *electoral fraud*;
4. **(D)** *high treason and terrorism*;
5. **(E)** *military classified information offence*.

The dependent variable (business cycle impact) will be the *GDP Dynamics* (*Real GDP per capita*). Thus the dependent variable will be quantitatively comparable to the independent variables.

Table 1 shows the results of the regression analysis:

Table 1

*Regression analysis of the impact of hybrid warfare on GDP*

|  | Synt. Coef. (SC) | Coef. (A) | Coef. (B) | Coef. (C) | Coef. (D) | Coef. (E) |
|---|---|---|---|---|---|---|
| **R** | 0.81582 | 0.81582 | 0.72921 | 0.33305 | 0.68904 | 0.54634 |
| **$R^2$** | 0.66557 | 0.66557 | 0.53175 | 0.11092 | 0.47478 | 0.29849 |
| **Adjusted $R^2$** | 0.62841 | 0.62841 | 0.47972 | 0.01213 | 0.41642 | 0.22055 |
| **Stand. Error** | 271.865 | 271.865 | 321.694 | 443.277 | 340.702 | 393.750 |
| **Sign. F** | **0.00219** | **0.00219** | **0.01088** | 0.31691 | **0.01901** | 3.82960 |
| **P-value (XVar.)** | 0.00219 | 0.00219 | 0.01088 | 0.31691 | 0.01901 | 0.08204 |

**Source:** Author's calculations.

The analysis covers a period of eleven years (2007 through 2017), which guarantees the reliability of the results. The empirical data is derived from the following sources: EUROSTAT (2018), NSI (2018), SJC (2018).

The synthetic coefficient (SC) is a combination of independent variables (A, B, C, D, E). It is used as an independent variable in the regression analysis.

Analysis of results. The correlation coefficient R varies between 0.81 and 0.68 and indicates a positive correlation between the dependent variable and the independent variables (SC, A, B, D). The coefficient of determination R2 varies between 0.66 and 0.47, which means that 66 to 47% of the changes of the dependent variable are caused by the independent variables (SC, A, B, D). The adjusted R2, as a measure of explanatory power, is high enough for the independent variables SC, A, B, D. Fisher's criterion has a standard significance level of 0.05. Since the values of the independent variables (SC, A, B, D) are below the significance threshold, the dependence between them and the resulting variable is statistically significant.

The same applies to the significance level of the P-value (0.05) of the XVar, which is statistically significant for the independent variables (SC, A, B, D). The absence of statistical significance of independent variables C and E in single linear regression does not undermine their explanatory power as components in the synthetic independent variable (SC).

The results of the regression analysis are summarized as:
(Equation 1)

**GDP Dynamics =** $f_{n1,n2 \ldots nx}$ **[(*public corruption crimes*) + (*cybercrimes*) + (*electoral fraud*) + (*high treason and terrorism*) + (*military classified information offence*)]** (1)
*(Real GDP per capita: n1, n2, ... nx)*

The results can be interpreted as follows:
Regarding the first objective of the study - to improve crisis forecasting and business cycle management, as well as contribute to the economic theory in this field - the regression analysis proved the following relationships:

A) Between the complex impact of the synthetic independent variable representing hybrid warfare on the dependent variable representing the business cycle;

B) Between the single impact of a some of the independent variables that could indicate the beginning of hybrid war on the dependent variable representing the business cycle.

The results for the studied period using empirical data for Bulgaria can be interpreted as follows. There is a significant relationship between the fluctuations of the Bulgarian business cycle and the change in the number of certain crime types represented as independent variables in the econometric model. Projections of these particular categories of crime can be used in forecasting the business cycle.

Regarding the second objective of the study – to use an economic approach to supplement the scientific knowledge in forecasting and management of hybrid warfare:

The results of the regression analysis show that it is possible to identify certain hybrid attacks and threats that are covert, i.e. do not involve armed conflicts. Thus, the economic approach can be used in the field of national security in addition to the other scientific approaches used for this purpose.

Regarding the second objective of the study, the results for the studied period using empirical data for Bulgaria can be interpreted as follows. There are indications for moderate use of covert hybrid warfare. However, these indications do not identify the type of the aggressor behind them (i.e. whether it is a neighbouring state or a non-state entity, such as a terrorist organization.)

**Conclusion**

The literature review shows that the objectives and the methods used in this study have not been used in other studies in this field. The econometric model could complement the economic toolkit for dealing with exogenous macroeconomic shocks with a new tool for managing hybrid warfare. The study could also contribute to a better interdisciplinary study of hybrid warfare by adding economic analysis and an econometric model that could also be used in the political, military and social sciences.

However, it should be borne in mind that the derived econometric model has been successfully tested in only one country (Bulgaria). One of the constraints of the survey is the lack of public data that could be sued in the model for other European countries. Therefore, this study could be continued in the future by including other researchers and securing the funding needed to obtain the necessary data for other countries in order to verify the model.

**References**

Abadie, A., & Gardeazabal, J. (2003). The economic costs of conflict: A case study of the Basque Country. *The American Economic Review*, *93*(1), pp. 113-132.

Bachmann, S. D., & Gunneriusson, H. (2015). Russia's Hybrid Warfare in the East: The Integral Nature of the Information Sphere. *Georgetown Journal of International Affairs*. (16) pp. 198-210.

Blattman, C., & Miguel, E. (2010). Civil War. *Journal of Economic Literature*, *48*(1), pp. 3-57.

Becker, Gary S. and Yona Rubinstein (2004). Fear and the Response to Terrorism: An Economic Analysis. Mimeo, University of Chicago.

Chen, S., Loayza, N. V., & Reynal-Querol, M. (2008). The aftermath of civil war. *The World Bank Economic Review*, *22*(1), pp. 63-85.

Cederman, L. E., Hug, S., Schädel, A., & Wucherpfennig, J. (2015). Territorial autonomy in the shadow of conflict: Too little, too late?. *American Political Science Review*, *109*(2), pp. 354-370.

Eckstein, Z., & Tsiddon, D. (2004). Macroeconomic consequences of terror: theory and the case of Israel. *Journal of Monetary Economics*, *51*(5), pp. 971-1002.

Guriev, S., & Melnikov, N. (2016). War, inflation, and social capital. *The American Economic Review*, *106*(5), pp. 230-235.

Hoffman, F. G. (2009). Hybrid threats: Reconceptualizing the evolving character of modern conflict. Strategic Forum, 240, pp.1–8.

Johnson, D. D., & MacKay, N. J. (2015). Fight the power: Lanchester's laws of combat in human evolution. *Evolution and Human Behavior*, *36*(2), pp. 152-163.

Lerer, Z., & Amram-Katz, S. (2011). The Sociology of Military Knowledge in the IDF: From 'Forging 'to' Deciphering'. *Israel Studies Review*, *26*(2), pp. 54-72.

Meierrieks, D., & Gries, T. (2013). Causality between terrorism and economic growth. *Journal of Peace Research*, *50*(1), pp. 91-104.

Reeves, S. R., & Barnsby, R. E. (2013). The New Griffin of War: Hybrid International Armed Conflicts. *Harvard International Review*, *34*(3), p. 16.

Sandler, T., & Enders, W. (2008). Economic consequences of terrorism in developed and developing countries. *Terrorism, economic development, and political openness*, p. *17*.

Wilmshurst, E. (Ed.). (2012). *International law and the classification of conflicts*. Oxford University Press. pp. 16-29.

EUROSTAT (2018). Real GDP per capita. Available at: https://ec.europa.eu/eurostat/data/database

NSI (2018). National Statistical Institute-Republic of Bulgaria, Demographic and social statistics, Justice and crime. Available at: http://www.nsi.bg/en/node/6210

SJC(2018). Supreme Judicial Council-Republic of Bulgaria, Judicial Statistics. Available at: https://portal.justice.bg/

# ECONOMIC ARCHIVE

**YEAR LXXII, BOOK 4 – 2019**

*CONTENTS*

# Economic Archive

# Requirements to be met when depositing articles for Narodnostopanski arhiv journal

**1. Number of article pages:** from 12 to 25 standard pages
**2. Deposit of journal articles:** one printout (on paper) and one in electronic form as attached file on E-mail: NSArhiv@uni-svishtov.bg
**3. Technical characteristics:**
- performance Word 2003 (minimum);
- size of page – A4, 29-31 lines and 60-65 characters on each line;
- line spacing 1,5 lines (At least 22 pt);
- font – Times New Roman 14 pt;
- margins – Top - 2.54 cm; Bottom - 2.54 cm; Left - 3.17 cm; Right - 3.17 cm;
- page numbering – bottom right;
- footer text – size 10 pt;
- graphs and figures – Word 2003 or Power Point.

**4. Layout:**
- title of article, name of author, academic position and academic degree – font Times New Roman, 14 pt, with capital letters Bold – centered;
- workplace, postal address, telephone and E-mail;
- abstract in Bulgarian up to 15 lines; key words – 3 to 5;
- JEL classification of publications on economic topics (http://ideas.repec.org/j/index.html);
- main body ( main text);
- tables, graphs and figures are software inserted in the text (they should allow linguistic corrections and translation in English). Numbers and text in them should be written with font Times New Roman 12 pt;
- formulas are inserted with Equation Editor.

**5. Rules for footnote:** When citing sources, authors should observe the requirements of **APA Style** at http://www.apastyle.org/ or at http://owl.english.purdue.edu/owl/resource/560/01/ or at http://www.calstatela.edu/library/guides/3apa.pdf .

Each author is responsible for promoting ideas, content and technical layout of the text.

**6**. **Manuscripts of lecturers without an academic** rank should be accompanied by a transcript of the minutes of the Department meeting at which the proposed paper was discussed.

From 1st of January 2017 the English language title of the journal is changed from "Narodnostopanski archiv" (transliterated from Bulgarian) to "Economic Archive".

Authors of papers published in Narodnostopanski arhiv journal are responsible for the authenticity of the materials.

From the Editorial Board

www.uni-svishtov.bg/NSArhiv